\documentclass[preprint]{revtex4-1}

\usepackage{amsmath,amsfonts,amssymb}
\usepackage{mathrsfs}
\usepackage{xparse}
\usepackage{physics}
\usepackage{booktabs}
\usepackage{graphicx}
\usepackage[utf8]{inputenc}
\usepackage[english]{babel}
\usepackage{hyperref}

\begin{document}
\title{Slow Light Frequency Reference Cavities -- Proof of Concept for Reducing the Frequency Sensitivity Due to Length Fluctuations}

\author{Sebastian P.\ Horvath}
\author{Chunyan Shi}
\author{David Gustavsson}
\author{Andreas Walther}
\author{Adam Kinos}
\author{Stefan Kr\"oll}
\author{Lars Rippe}
\email{physics@rippe.se}
\affiliation{Department of Physics, Lund University, P.O. Box 118, SE-22100 Lund, Sweden.}

\date{\today}

\begin{abstract}
Length changes due to thermo-mechanical noise originating from, for example, Brownian motion are a key limiting factor of present day state-of-the-art laser frequency stabilization using Fabry-Pérot cavities. We present a laser-frequency stabilization concept using an optical cavity with a strong slow-light effect to reduce the impact of cavity length changes on the frequency stability. The resulting noise-reduction factor is proportional to the ratio between the light phase and group velocities in the highly dispersive cavity spacer. We experimentally demonstrate a proof-of-principle implementation of this laser-frequency stabilization technique using a rare-earth doped crystalline cavity spacer in conjunction with semi-permanent spectral tailoring to achieve precise control of the dispersive properties of the cavity. Compared to the same setup in the absence of the slow-light effect a reduction in frequency sensitivity of four orders of magnitude was achieved.
\end{abstract}

\maketitle

\section{Introduction}

Precision frequency stabilized lasers are a fundamental component of modern metrology. They serve as a local oscillator in optical atomic clocks~\cite{ludlow2015}, and are at the heart of interferometer based tests of fundamental physics, such as gravitational wave detection~\cite{ligo2016} and tests of relativity~\cite{chou2010,eisele2009}. State-of-the-art frequency stabilization is performed using Fabry-P\'erot (FP) cavities in conjunction with the Pound-Drever-Hall (PDH) locking technique. In order to achieve good performance, thermo-mechanical isolation of the cavity, along with a careful material selection, is imperative.  Typical cavities, with a length on the order of $10\;\mathrm{cm}$, have a fractional stability of $10^{-15}$~\cite{ludlow2007,webster2008,chen2014}. 

Present day leading stabilities are achieved by utilizing long cavities ($\sim50\;\mathrm{cm}$), with a fractional stability of $8 \times 10^{-17}$~\cite{haefner2015}, or cryogenically cooling single-crystal silicon cavities, with a fractional stability of $4 \times 10^{-17}$~\cite{matei2017}. Such designs have been shown to operate at the frequency stability limit imposed by the thermo-mechanical noise in the mirror coatings~\cite{matei2017,numata2004}. In addition to utilizing long cavities~\cite{haefner2015,jiang2011}, or operating single-crystal cavities at cryogenic temperatures~\cite{seel1997,matei2017,kessler2012}, investigations to reduce the thermo-mechanical noise originating from the mirror coatings has focused on increasing the size of the optical mode~\cite{amairi2013}, as well as utilizing single-crystal mirrors with low dielectric losses~\cite{cole2013,brueckner2010}.

While the fractional stability has been considerably improved over the last decade by refining cavity designs, the thermo-mechanical noise limit of FP cavities has prompted investigations into alternative stabilization techniques including cavity quantum-electrodynamic systems~\cite{bishof2013,christensen2015,norcia2016} and locking to spectral holes in rare-earth ion doped crystals~\cite{sellin1999,julsgaard2007,thorpe2011,chen2011,leibrandt2013,cook2015,gobron2017,galland2020}. 

The slow-light effect consists of a reduction of the group velocity of light in a medium due to a sharp gradient in the dispersion. This effect has been successfully utilized in conjunction with optical cavities to demonstrate a reduced sensitivity of the cavity resonance to length changes as well as a narrowing of the cavity linewidth \cite{pati2008,yablon2017,sabooni2013}. In this work we demonstrate the locking of a laser to a slow-light cavity, harnessing both the reduced sensitivity to length changes as well as the cavity linewidth narrowing. 

\begin{figure}[tb!]
\centering
\includegraphics[width=\textwidth]{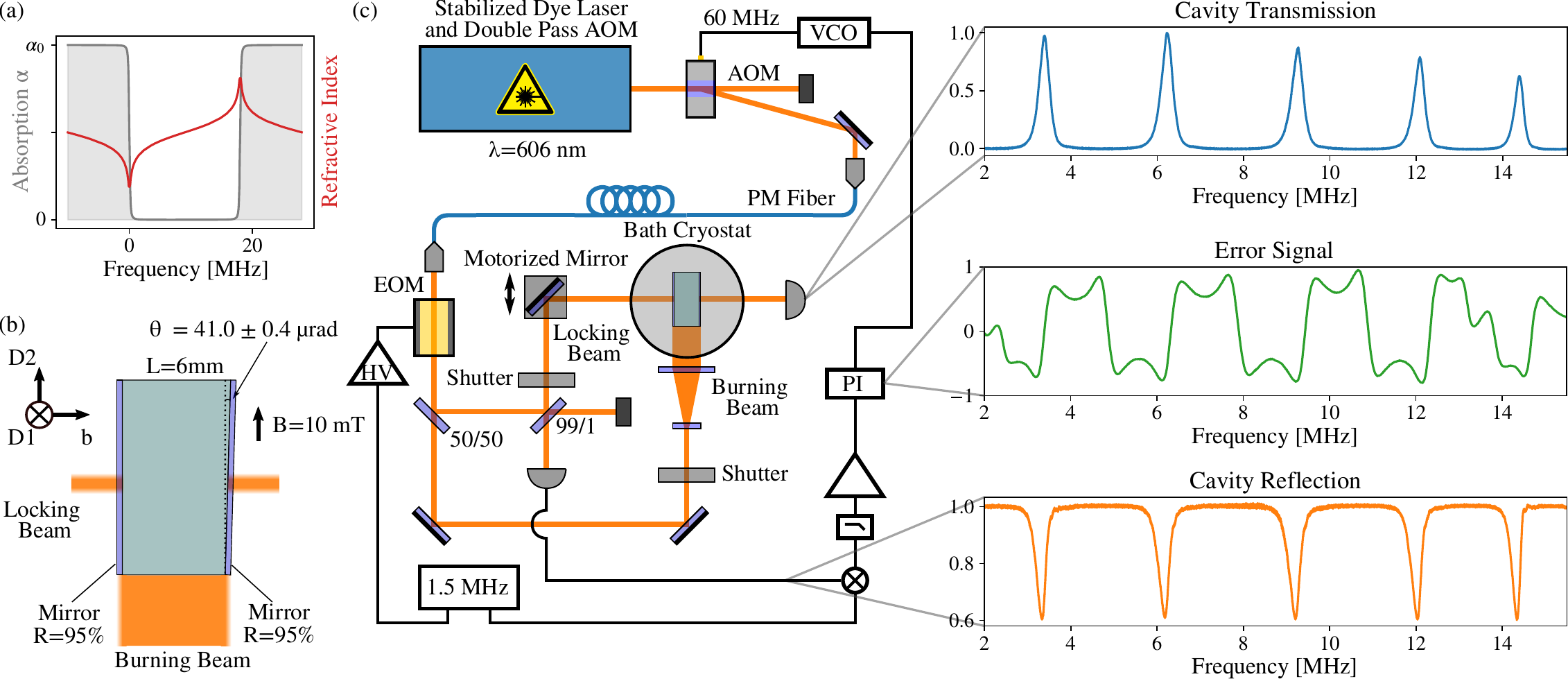}
\caption{(a) Schematic of a spectral hole and the resulting rapidly varying dispersion across this frequency region. (b) Schematic of the crystal geometry, showing the burning beam perpendicular to the locking beam. (c) The experimental layout of the core laser-locking setup, with representative signals observed at key locations in the locking loop. The EOM modulation was disabled to record the cavity transmission and reflection signals. The steep gradient visible in the error signal at frequencies where there is no peak in the reflection signal originates from the large modulation used ($1.5\;\mathrm{MHz}$, which is almost half of the cavity mode spacing). For a detailed description of these signals in the context of PDH locking, see Ref.~\cite{black2000}.}
\label{fig:exp}
\end{figure}

The cavity is constructed using a rare-earth ion doped crystal with mirror coatings directly deposited on the crystal surfaces. By spectral tailoring, a narrow semi-permanent frequency transmission window is burned into the inhomogeneously broadened rare-earth ion ensemble. The narrow transmission window leads to a gradient of the index of refraction across the spectral window, which reduces the speed of light in the cavity by up to four orders of magnitude (see figure~\ref{fig:exp}(a)). Such an effect has been previously demonstrated to lead to a narrowing of the cavity-linewidth by a factor comparable to the reduction of the speed of light~\cite{sabooni2013}. 

The slow-light effect increases the cavity round-trip time, which, heuristically, can be thought of as an increase of the effective cavity length from $6\;\mathrm{mm}$ to $60\;\mathrm{m}$. Therefore, this technique enables designs with a very large effective length without the mechanical degradation associated with a physically longer cavity. More concretely, assuming an inhomogeneously broadened ensemble with an absorption $\alpha_0$ and a square spectral transmission window, as indicated in figure~\ref{fig:exp}, the cavity free-spectral range (FSR) has the form
\begin{equation}
  \nu_{\text{FSR}} = \frac{v_g}{2L} = \frac{1}{2L} \frac{\pi^2 \Gamma}{\alpha_0} = \frac{c}{2L} \frac{1}{n+ \nu \frac{dn}{d\nu}}.
  \label{eqn:sl_cav_modes}
\end{equation}
Here, $v_g$ is the group velocity of light, $L$ is the cavity length, $\Gamma$ is the transparency window width, $c$ is the speed of light, $n$ is the refractive index, and $\nu$ the frequency of the light; a derivation of this relation is provided in appendix~\ref{sec:group_vel}. Note that the above equation is only valid at the center of the transparency window; nevertheless, since the dispersive gradient is close to constant across a large portion of the transparency window, equation~\eqref{eqn:sl_cav_modes} is a good approximation for all cavity modes not adjacent to the edges. 

The slow-light effect not only reduces the cavity FSR, but equally scales the cavity linewidth $\delta \nu_{\text{cav}}$. Assuming a high-finesse cavity with only small mirror losses~\cite{siegman1986}, one finds
\begin{equation}
  \delta \nu_{\text{cav}} = \frac{\nu_{\text{FSR}}}{\mathcal{F}} = \frac{v_g}{2L} \left(\frac{T_1 + T_2 + 2\alpha_c L}{2 \pi} \right).
  \label{eqn:sl_cav_fwhm}
\end{equation}
Here, $\mathcal{F}$ is the cavity finesse, $\alpha_c$ is the residual absorption at the center of the transparency window due to off-resonant excitation, while $T_1$ and $T_2$ are the transmission of the first and second mirror. From equations~\eqref{eqn:sl_cav_modes} and \eqref{eqn:sl_cav_fwhm} it is apparent that the finesse $\mathcal{F}$ is independent of the slow-light effect. 

In addition to a narrowing of the cavity modes, the slow-light effect also results in an insensitivity to cavity-length changes. By considering the relative frequency shift due to a cavity length change, one obtains
\begin{equation}
  \frac{d \nu}{\nu} = -\frac{d L}{L} \frac{v_g}{c}.
  \label{eqn:freq_noise}
\end{equation}
The same scaling with respect to $v_g$ is true for the frequency noise amplitude $\sqrt{S_{\nu}}$ resulting from a corresponding displacement noise amplitude. As a consequence, all optical length-change noise sources, including from the mirror coatings, are reduced by a factor corresponding to the group velocity divided by the speed of light.

\section{Experimental results}

The proof-of-principle implementation utilized a Pr$^{3+}$:Y$_2$SiO$_5$ crystal as a spacer material for an FP cavity with mirror coatings applied directly to two, almost parallel, flat crystal surfaces. While a better ultimate stability could be achieved by utilizing a material with a narrower optical homogeneous linewdith and slower hyperfine relaxation rate, the present demonstration of this locking technique was implemented in Pr$^{3+}$:Y$_2$SiO$_5$ due to the well established spectral-tailoring routines available for this material~\cite{nilsson2004}. An optical transmission window was burned at $494.723\;\mathrm{THz}$, on resonance with the $^3$H$_4 \to ^1$D$_2$ transition of Pr$^{3+}$ ions. The geometry utilized for spectral 
holeburning with respect to the crystal orientation and the mirror coatings is shown in figure~\ref{fig:exp}(b).

\begin{figure}[tb!]
\centering
\includegraphics[width=\textwidth]{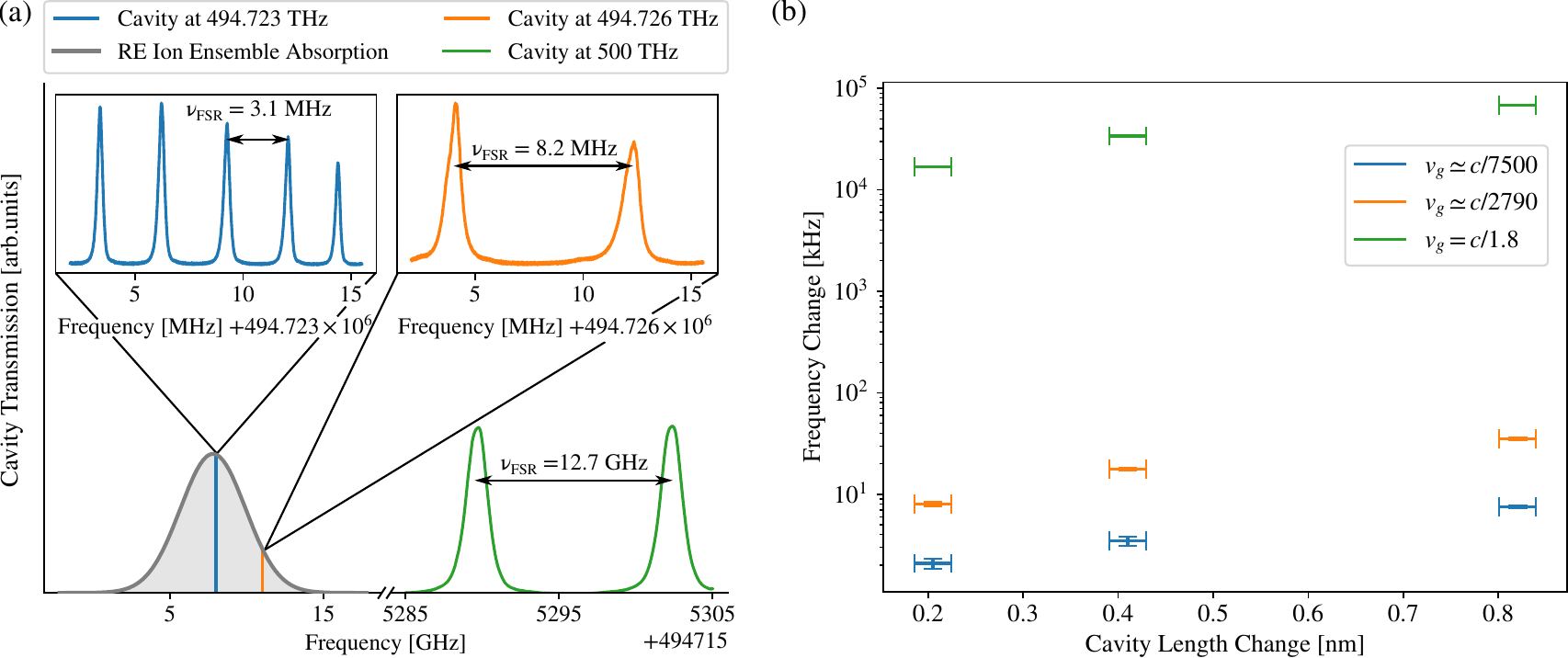}
\caption{(a) Representative cavity modes observed at three distinct frequencies relative to the inhomogeneously broadened rare-earth ion absorption profile. Colors designate the cavity frequency: blue is at maximum absorption, orange is detuned from the maximum absorption by 3 GHz, and green is far away from the absorption peak, detuned by $\simeq 5.3\;\mathrm{THz}$, which is sufficient to ensure that the slow-light effect is negligible. The inhomogeneous line, shown in grey, has an FWHM of $\simeq 5\;\mathrm{GHz}$, determined by an absorption measurement in a portion of the crystal that was not mirror coated. The frequencies of the two insets relative to the inhomogeneous profile are indicated using the blue and orange vertical bars. (b) The frequency sensitivity to cavity length changes at the previously enumerated cavity frequencies (see figure~\ref{fig:exp} for the experimental configuration). The blue and orange measurements were performed with the laser locked to the slow-light cavity, while the green measurements were performed by directly measuring the shift of the cavity modes due to the induced length changes. See text for further details. The indicated group velocities were determined from the change of the cavity FSR. The frequency sensitivity can be seen to scale with $v_g$; any deviation from an exact correspondence is attributed to a deviation from an optimally square spectral transparency window.}
\label{fig:cav_modes}
\end{figure}

Figure~\ref{fig:exp}(a) schematically shows the spectral transmission window and the corresponding change in refractive index. The gradient of the dispersion leads to, in accordance with equations~\eqref{eqn:sl_cav_modes} and \eqref{eqn:sl_cav_fwhm}, a reduction in both the cavity mode spacing and the linewidth $\delta \nu_{\text{cav}}$. The group velocity of light propagating through the transparency window scales inversely with the absorption, $\alpha_0$, outside the transparency window; therefore, the cavity mode linewidth is strongly dependent on the frequency of the spectral hole with respect to the Pr$^{3+}$ inhomogeneous line. This is illustrated in figure~\ref{fig:cav_modes}(a), which shows the transmission cavity modes for three scenarios with different values for $\alpha_0$. Specifically, a spectral transmission window with a width of $18\;\mathrm{MHz}$ was burned at the center of the Pr$^{3+}$ ensemble, as well as detuned from the center by $3\;\mathrm{GHz}$. This yielded a full width at half maximum (FWHM) cavity mode linewidth of $275 \pm 2\;\mathrm{kHz}$ and $745 \pm 14\;\mathrm{kHz}$, respectively. In contrast, the cavity linewidth obtained when the laser was detuned by $\simeq 5.3\;\mathrm{THz}$, where we expect any absorption from Pr$^{3+}$ ions to be negligible, was $1317 \pm 2\;\mathrm{MHz}$.

In order to test the reduction in frequency sensitivity to cavity length changes, and consequently corroborate equation~\eqref{eqn:freq_noise}, a small relative angle between the mirror coated sides of the Pr$^{3+}$:Y$_2$SiO$_5$ crystal was exploited. This angle is schematically shown in figure~\ref{fig:exp}(b), and amounted to a relative slope of $41.0 \pm 0.4\;\mathrm{\mu rad}$. The cavity input beam was equipped with a motorized mirror mount, such that, when the input beam was translated along the slope, the cavity length was changed by a calibrated amount. Consequently, by translating the input beam while monitoring the resulting cavity frequency shift for the three distinct slow-light regimes described above (see figure~\ref{fig:cav_modes}(a)), it was possible to probe the impact of the slow-light effect on the frequency shift as a function of length change. More specifically, in order to determine the cavity mode frequency shift when the laser was on resonance with the Pr$^{3+}$:Y$_2$SiO$_5$ inhomogeneous profile, as well as when it was detuned by $3\;\mathrm{GHz}$, the PDH technique was employed~\cite{black2000}. A pre-stabilized laser was passed through an acousto-optic modulator (AOM) followed by a PDH lock to the slow-light cavity, as shown in figure~\ref{fig:exp}(c); a detailed experimental description is provided in appendix~\ref{sec:exp}. The PDH feedback signal was applied to the AOM to create a closed loop. Monitoring the feedback signal thus provided a high-resolution measurement of the stability and drift attainable by the slow-light cavity. To ensure the locking beam propagated within a spectral hole throughout the translation, a spatial region of the crystal surrounding the locking beam was spectrally tailored using a large burning beam perpendicular to the locking beam.  

This measurement was repeated for three different translation step sizes. For each step size, the experiment was cycled 10 times, and the resulting frequency shift averaged. By performing this measurement both at the center of the inhomogeneous Pr$^{3+}$ ensemble, as well as detuned by $3\;\mathrm{GHz}$, the dependence of the induced frequency change could be determined with respect to the slow-light effect. These measurements are shown in figure~\ref{fig:cav_modes}(b) for, respectively, the group velocities of $v_g \simeq c/7500$ and $v_g \simeq c/2790$. The frequency shift for each step size was also determined in the absence of the slow-light effect by detuning the laser by $\simeq 5.3\;\mathrm{THz}$. However, due to the large increase in the cavity FSR, bandwidth limitations precluded the use of the PDH method outlined above. The frequency change was instead determined by directly measuring the cavity modes in transmission before and after each translation step (see appendix~\ref{sec:data_analysis} for further details). These measurements are also included in figure~\ref{fig:cav_modes}(b), denoted by the speed of light in Y$_2$SiO$_5$ of $v_g = c/1.8$. We note that this last measurement was also used to calibrate the angle of the crystal. 
\begin{figure}[tb!]
\centering
\includegraphics[width=0.6\columnwidth]{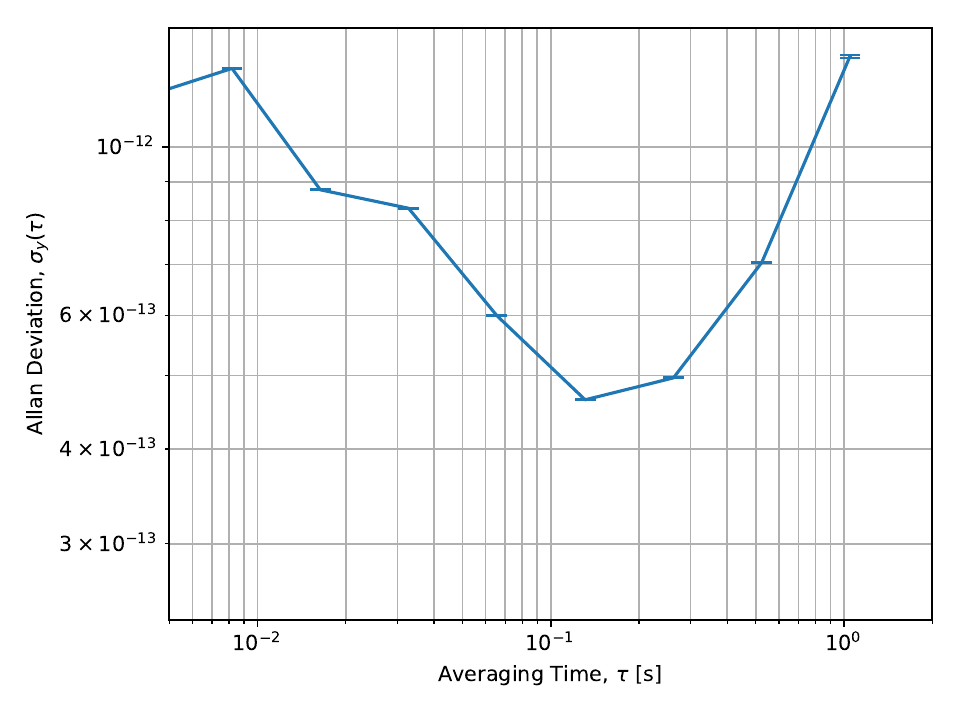}
\caption{Overlapped Allan deviation with linear drift subtracted \cite{riley2008}. The noise floor was found to be $\sigma_y(\tau) = 4.6 \times 10^{-13}$ at an averaging time of $130\;\mathrm{ms}$. }
\label{fig:oadev}
\end{figure}

This measurement shows that the frequency shift due to length changes scales proportionally with the group velocity, and achieves a reduction in length-change sensitivity of almost four orders of magnitude at the center of the Pr$^{3+}$ ensemble. From this measurement, the minimal drift rate of our locking setup was found to be $0.55 \pm 0.06\;\mathrm{kHz}/s$. The principal limiting factors for these experiments, as will be discussed in detail below, originated from two distinct sources of atomic drift. Furthermore, the locking loop included a single-mode fiber which did not include any Doppler noise cancellation.  Figure~\ref{fig:oadev} shows the overlapped Allan deviation after linear drift subtraction \cite{riley2008}, with a noise floor of $\sigma_y(\tau) = 4.6 \times 10^{-13}$ at an averaging time of $130\;\mathrm{ms}$. One possible explanation for this limit could be the polarization-maintaining fiber that is part of the locking loop (see figure~\ref{fig:exp}) which for a 25 m fiber has been shown to lead to frequency broadening up to $\sim 1.2$ kHz~\cite{ma1994}.  
\begin{figure}[tb!]
\centering
\includegraphics[width=\textwidth]{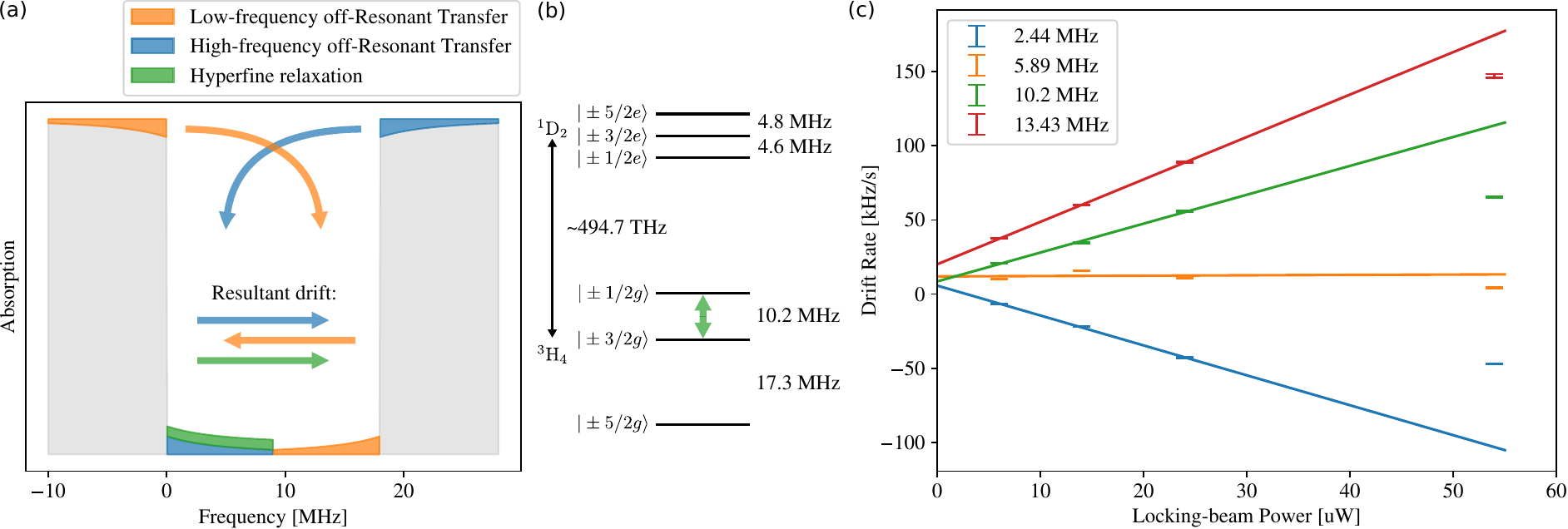}
\caption{(a) Changes in absorption in the spectral transmission window during locking and the resulting drift. Blue and orange shaded regions indicate off-resonant excitation and subsequent relaxation into the transmission window. The green shaded region is the result of direct hyperfine level-cross relaxation and occurs primarily at the low frequency side of the transmission window. (b) The energy level diagram of the Pr$^{3+}$:Y$_2$SiO$_5$ ground and excited state; the green arrow indicates the most likely hyperfine cross-relaxation pathway. (c) The locked laser drift rate, relative to the stabilized reference laser, when locked at different frequencies with respect to the center of the spectral transmission window. The non-linear effect for the higher locking intensity data is attributed to a re-burning of the spectral transmission window by the locking beam.}
\label{fig:drift_wrt_power}
\end{figure}

To further analyze the drift, the laser was locked to different cavity modes across the transmission window. This revealed two unique sources for the frequency drift: off-resonant excitation, and hyperfine cross-relaxation. Specifically, the circulating light in the cavity will off-resonantly excite ions at the high and low frequency edges of the spectral transmission window, which then have a non-negligible probability of relaxing to a different hyperfine level. Ions excited from the high-energy edge in figure~\ref{fig:drift_wrt_power}(a) predominantly reside in the $\ket{\pm 5/2g}$ state. When excited they can decay to the $\ket{\pm1/2g}$ or $\ket{\pm3/2g}$ states. This transfer can, effectively, be considered as a shift of the spectral transmission window. As a consequence, the dispersion curve within the spectral transmission window translates to higher frequencies, which also shifts the cavity modes towards higher frequencies. This process is counteracted by the off-resonant excitation of ions at the low energy edge of the transmission window in figure~\ref{fig:drift_wrt_power}(a). Here, ions occupy the $\ket{\pm1/2g}$ state (see figure~\ref{fig:drift_wrt_power}(b)), but may, when they are excited, fall down in either the $\ket{\pm3/2g}$ or $\ket{\pm5/2g}$ states, effectively leading to a negative shift of the cavity modes. These two cases of positive and negative frequency drift are indicated by the blue and orange shaded regions in figure~\ref{fig:drift_wrt_power}(a).  

The second source of drift stems from the hyperfine relaxation between the $\ket{\pm 1/2g}$ state and the $\ket{\pm 3/2g}$ state. Relaxation from $\ket{\pm1/2g}$ to $\ket{\pm3/2g}$ results in absorption at low frequencies of the spectral transmission window; consequently, the cavity modes drift to higher frequencies. This is indicated by the shaded green region in figure~\ref{fig:drift_wrt_power}(a). Figure~\ref{fig:drift_wrt_power}(c) shows the drift with respect to different locking beam intensities. The hyperfine relaxation rate is independent of the locking beam intensity, and therefore leads to a constant positive offset in the drift rate.  

The data summarized in figure~\ref{fig:drift_wrt_power}(c) establishes that, in general, the drift due to off-resonant excitation depends on the locking beam intensity and the frequency position of the cavity mode relative to the transparency window. At high locking intensities, the drift becomes non-linear due to a re-burning of the spectral transmission window by the locking beam. The only point where there is no clear power dependence is for the $5.89\;\mathrm{MHz}$ locking point. Here the constant drift is solely due to hyperfine cross-relaxation, which demonstrates that the drift due to the low and high frequency off-resonant excitation can be cancelled. The drift may be mitigated by carefully positioning the cavity mode at the center of the transparency window and choosing a material with negligible hyperfine cross-relaxation during practical locking timescales. This can be realized by utilizing Eu$^{3+}$:Y$_2$SiO$_5$, for which persistent spectral holes have been observed for up to $50\;\mathrm{days}$ after creation~\cite{oswald2018}. The relative position of the cavity mode within the transparency window can be fine tuned by moving the window within the inhomogeneous profile. 

While this demonstrates that it is possible to mitigate the observed drift with a carefully selected locking point and material selection, off resonant excitation will nevertheless slowly degrade the spectral structure and therefore impair the attainable slow-light effect after extended locking periods. The magnitude of this effect is determined by the optical power circulating in the cavity, and therefore requires a careful selection of locking parameters. This is discussed in detail below, where we perform an estimate of the attainable stability for an optimized slow-light cavity. Finally, we note that even when locked to a cavity mode that cancels drift due to off-resonant excitation, any absolute drift in the resonance frequency of the ions is nevertheless a source of noise for a locked laser. In particular, ambient pressure and temperature changes have been determined to impact spectral-hole locking techniques~\cite{thorpe2013}, and are likely also key limiting sources of drift for the presented locking technique. However, to our knowledge, this ultimate atomic limit has yet to be achieved by any of the rare-earth ion based laser frequency stabilization techniques presently under investigation~\cite{leibrandt2013,galland2020}.

\section{Discussion \label{sec:discussion}}

We now discuss several factors which impact the performance of a slow-light cavity. In order to describe these factors, it is necessary develop several prerequisites which are interesting in their own right. The integrated off-resonant absorption at the center of a spectral transmission window can be approximately expressed as
\begin{equation}
 \alpha_c \simeq  \frac{2}{\pi}\frac{\Gamma_h}{\Gamma} \alpha_0.
 \label{eqn:off_res_abs}
\end{equation}
Here, $\alpha_0$ is the absorption outside the transmission window, $\Gamma_h$ is the homogeneous linewidth of the ions, and $\Gamma$ is the transmission window width~\cite{nilsson2005}. Substituting equation~\eqref{eqn:off_res_abs} into equation~\eqref{eqn:sl_cav_fwhm}, and using the expression for $v_g$ in terms of $\alpha_0$ and $\Gamma$ given by equation~\eqref{eqn:sl_cav_modes}, it is straightforward to show that the cavity linewidth takes the form
\begin{equation}
  \delta \nu_{\text{cav}} \simeq \Gamma_h + \frac{\pi \Gamma (T_1 + T_2)}{4 \alpha_0 L}.
  \label{eqn:cav_fwhm_homo}
\end{equation}
Considering, for simplicity, the case of an impedance matched cavity, that is, imposing the requirement that $T_1 = 2 \alpha_c L$, and an output mirror transmission approaching zero, the above equation reduces to 
\begin{equation}
  \delta \nu_{\text{cav}} \simeq 2 \Gamma_h.
  \label{eqn:cav_fwhm_T2lim}
\end{equation}
Interestingly, in this scenario the linewidth is, to first order, independent of the cavity length, absorption coefficient, and the transmission window width (provided the cavity mirrors are adjusted to maintain impedance matching); the linewidth is entirely determined by the homogeneous linewidth of the active resonant absorber ions. 

Relating the cavity mode linewidth to the Allan deviation, one obtains
\begin{equation}
  \sigma_y(\tau) = \frac{1}{\sqrt{2\tau}} \frac{1}{4} \sqrt{\frac{h \nu}{P_c}} \delta \nu_{\text{cav}},
  \label{eqn:allen_dev}
\end{equation}
where $P_c$ is the shot-noise limited locking power, $\tau$ the locking time, $\nu$ the locking frequency, and $h$ corresponds to Planck's constant (see appendix \ref{sec:shotnoise_lim} for a derivation). Consequently, this equation can be used to determine the requisite optical power for a given Allan deviation. Substitution of equation \eqref{eqn:cav_fwhm_T2lim} further establishes that the material homogeneous linewidth directly fixes the necessary circulating optical power, and, as discussed above, the impact of off-resonant excitation. Considering for concreteness Eu$^{3+}$:Y$_2$SiO$_5$, the homogeneous linewidth is $200\;\mathrm{Hz}$~\cite{equall1994} compared to $2.5\;\mathrm{kHz}$ for Pr$^{3+}$:Y$_2$SiO$_5$~\cite{equall1995}. From equation \eqref{eqn:cav_fwhm_T2lim} it can be seen that this would allow for a cavity mode width of $1\;\mathrm{kHz}$, even when including a reasonable margin for experimental imperfections. Substituting this into equation \eqref{eqn:allen_dev}, and setting a target Allan variance of $1\times10^{-17}$ for a 1 second measurement time, one obtains $\approx 0.4\;\mathrm{nW}$ for the shot noise limited locking power. 
In appendix~\ref{sec:lockt} we evaluate the rate of off-resonant excitation resulting from a given circulating locking power and the impact on the attainable locking time. For a power of $\approx 0.4\;\mathrm{nW}$ this yields a lower bound of more than two hours. While there may be additional unforeseen limits that impact the ultimately attainable frequency stability, a determination of this limit requires an experimental investigation of the demonstrated locking technique using a narrow homogeneous linewidth material such as Eu$^{3+}$:Y$_2$SiO$_5$.  

A corollary of equation~\eqref{eqn:cav_fwhm_T2lim} is that for an impedance matched cavity the narrowest achievable cavity linewidth is not affected by the slow-light effect. Intuitively this can be understood as follows: by narrowing the transparency window width $\Gamma$ the group velocity can be reduced, and in accordance with equation~\eqref{eqn:sl_cav_fwhm}, the cavity linewidth narrows. However, this leads to an increase in the off-resonant absorption $\alpha_c$ at the center of the transparency window, which, in order to maintain impedance matching, requires an increase of the input mirror transmission $T_1$. While this might suggest that the narrowest transparency window achievable for a given material $\Gamma_h$ is optimal in order to maximize the reduction in length change sensitivity, such an optimization impairs the longest attainable locking period. In particular, the off-resonant excitation width extending into the ions adjacent in frequency to the spectral transparency window scales with the width $\Gamma$. Specifically, 50\% of the absorption at the center of the spectral window comes from the atoms in the intervals $[-\Gamma, -\Gamma/2]$ and $[\Gamma/2, \Gamma]$ (see appendix~\ref{sec:offresex} for further details). Thus, the width $\Gamma$ determines the number of ions interrogated by the circulating light of the cavity, and consequently, the number of ions over which any deterioration of the spectral structure is averaged. Broadening the transparency window width increases the time that the spectral structure can be maintained, and, consequently, the amount of time that the slow-light cavity can be locked. 

At its core, this trade-off between the optical power used to interrogate the atomic ensemble, spread over the finite spectral width of the ensemble, is true for all rare-earth ion based laser stabilization techniques. For spectral-hole locking, the equivalent averaging takes place when one performs interleaved locking to a large number of individual spectral holes~\cite{thorpe2011}; however, the total number of interrogated ions for implementations realized to date is considerably fewer than for the technique presented here. 

An additional potential noise source for the presented locking technique that is not an issue for conventional FP cavities are changes of the refractive index of the cavity spacer material due to environmental variations in temperature and pressure. Noise of this type is also reduced by the slow-light effect according to 
\begin{equation}
	\frac{d \nu}{\nu} = -\frac{d n}{n} \frac{v_g}{c},
\end{equation}
with $n$ the refractive index of the spacer. A detailed analysis of this is beyond the scope of the present paper, and the order of magnitude where this will become significant in a material such as Eu$^{3+}$:Y$_2$SiO$_5$ is yet to be determined. Ultimately, like with any locking setup, there will be a wide range of environmental factors that impair the stability, such as temperature, pressure, and vibration. The impact of some of these will be reduced by the slow-light effect similar to what has been demonstrated here for cavity length changes, although this will not hold true universally for all noise sources. A detailed noise analysis of this locking scheme will be reported in a subsequent work.

\section{Conclusion}

We have presented a laser frequency stabilization technique that utilizes the slow-light effect within an FP cavity to reduce the sensitivity of the resonance frequencies to cavity length changes. Implementing such a locking system using a single-crystal Pr$^{3+}$:Y$_2$SiO$_5$ cavity spacer achieved a reduction in length change sensitivity by four orders of magnitude. The concept validation of this system achieved a drift rate of $0.55 \pm 0.06\;\mathrm{kHz/s}$ with a noise floor of $\sigma_y(\tau) = 4.6 \times 10^{-13}$ for a $6\;\mathrm{mm}$ long cavity. We have theoretically demonstrated that a different material with narrower homogeneous linewidth and lower hyperfine cross-relaxation should significantly improve these values, and the case of Eu$^{3+}$:Y$_2$SiO$_5$, which entails more complex spectral tailoring, is presently under active investigation.

\section{Acknowledgements}

The authors would like to thank Ulf Söderlund and Leif Johansson for their help with dicing the Pr$^{3+}$:Y$_2$SiO$_5$ crystal. This research was supported by the Swedish Research Council (no. 2016-05121, no. 2015-03989, no. 2016-04375, and 2019-04949), the Knut and Alice Wallenberg Foundation (KAW 2016.0081), the Wallenberg Center for Quantum Technology (WACQT) funded by The Knut and Alice Wallenberg Foundation (KAW 2017.0449), the Royal Physiographic Society in Lund, and the European Union FETFLAG program, Grant No.820391 (SQUARE) (2017.0449).

%
%
%
%
%

\appendix

\section{Experimental details \label{sec:exp}}

The laser used was a Coherent 699-21 ring-dye laser, with a custom intra-cavity EOM, stabilized to an external ULE cavity. The linewidth is estimated to be on the order of a few tens of $\mathrm{Hz}$, with a drift rate below $1\;\mathrm{Hz/s}$~\cite{alnis2008}. Prior to injection of the light into the locking loop, arbitrary pulse shaping in amplitude, frequency, and phase, was performed with a double-pass AOM. Locking was performed using the PDH technique, with a modulation frequency of $1.5\;\mathrm{MHz}$. To verify that a tight lock was achieved, the pre-stabilized laser was stepped through a series of frequency changes while the feedback loop was monitored. The sample was cooled to $2.1\;\mathrm{K}$ in an Oxford Instruments bath cryostat. The Y$_2$SiO$_5$ crystal has yttrium atoms replaced with praseodymium at a $500\;\mathrm{ppm}$ level; the sample is a cylinder of diameter $10\;\mathrm{mm}$ and a height of $6\;\mathrm{mm}$ (purchased from Scientific Materials). One side of the cylinder has been cut and polished to allow for the perpendicular burn geometry indicated in figure~\ref{fig:exp}(c). The polarization of both the burn and probe light was rotated to be parallel to the $D_2$ optical extinction axis of Y$_2$SiO$_5$, which has an absorption depth of $47 \pm 5\;\mathrm{cm}^{-1}$ \cite{sun2006}. The width of the spectral transmission window used throughout was $18\;\mathrm{MHz}$.

All measurements, with the exception of the drift rate with respect to the locking point, were performed with a magnetic field of $10\;\mathrm{mT}$ along the crystal $D_2$ axis, which enhanced the lifetime of the ground-state hyperfine levels. Moreover the drift rate data was taken in a configuration without the perpendicular burn-beam geometry; consequently burning was performed through the mirror coating. While this impaired the absolute drift performance, it still demonstrated the trends of the drift vs intensity when locking to different cavity modes.

In order to perform the beam-translation experiment, and to achieve optimal burning, the perpendicular burn geometry was utilized. The burn beam was shaped with cylindrical lenses and blocked with a mechanical shutter during locking. To avoid standing waves during burning, a second shutter blocked the locking beam during the burn operation. Additionally, for the measurements of the smallest drift rate, as well as the noise floor, the burning beam was dithered with a piezo driven mirror mount with a frequency of $10\;\mathrm{Hz}$, and a lens tissue paper inserted into the beam path before the cryostat window. By making the burning beam more diffuse, this ensured optimal burning of the entire length of the crystal (including close to the mirror coated sides), and mitigated residual drift from unburned spatial regions of the crystal because of (partial) standing waves arising from reflections of the parallel crystal surfaces. The error signal was amplified and $30\;\mathrm{kHz}$ low-pass filtered prior to locking with a Vescent D2-125.

\section{Group-velocity in spectral transparency window \label{sec:group_vel}}

To derive the group velocity for a pulse of light travelling through a rectangular spectral transmission window we follow the method suggested by Chanelière~\cite{chaneliere2021, lauro2009}. We note that, for a weakly-absorbing host with refractive index $n_0$, doped by an absorbing species, the imaginary component of the first order susceptibility reads~\cite{saleh2007}
\begin{equation}
  \chi''(\omega) = - \frac{n_0 G(\omega)}{k_0}.
  \label{eqn:chipp}
\end{equation}
Here, $G(\omega)$ is spectral absorption lineshape, $k_0 = \omega_0/c$, $\omega_0$ is the central angular frequency of the inhomogeneous absorption profile, which is assumed to be coincident with the center of the spectral transparency window, and $c$ is the vacuum speed of light. By application of the Kramers-Kroniger relation, the real part of the first order susceptibility can be evaluated from equation~\eqref{eqn:chipp} using
\begin{equation}
  \chi'(\omega) = \frac{1}{\pi} \text{ P.V.}\int_{-\infty}^\infty \frac{\chi''(\omega')}{\omega - \omega'} d\omega',
  \label{eqn:chip}
\end{equation}
where P.V.\ denotes the Cauchy principal value. For a square spectral transmission window of width $\Gamma$ (in units of frequency), which is much narrower than the inhomogeneous absorption profile, the transparency window can be approximated by
\begin{equation}
  G'(\omega) = \alpha_0(1 - \Pi(\omega)).
  \label{eqn:rect_abs}
\end{equation}
Here, $\alpha_0$ is the peak absorption outside the transparency window and $\Pi(\omega)$ is the rectangular function defined by
\begin{equation}
  \Pi(\omega) = 
  \begin{cases}
    0, & | \omega -\omega_0 | > \pi \Gamma \\
    1/2, & | \omega -\omega_0 | = \pi \Gamma \\
    1, & | \omega -\omega_0 | < \pi \Gamma \\ 
  \end{cases}
  \label{eqn:rect}
\end{equation}
When directly evaluated, the integral equation~\eqref{eqn:chip} is manifestly divergent. This divergence could be alleviated by convolving $G'(\omega)$ with an inhomogeneous absorption lineshape such as a Gaussian; however, the resulting integral is difficult to evaluate analytically. To circumvent this, we note that the integral equation~\eqref{eqn:chip} for a flat distribution can be shown to be zero (that is, for a perfectly flat absorption profile, the dispersion is zero). Consequently,  
\begin{equation}
  \int_{-\infty}^\infty G'(\omega) d\omega = \int_{-\infty}^\infty -G(\omega) d\omega,
\end{equation}
where $G(\omega) = \alpha_0 \Pi(\omega)$.  

Substituting the appropriate terms into equation~\eqref{eqn:chip}, one obtains,
\begin{align}
  \chi' &= \frac{1}{\pi} \frac{\alpha_0 n_0}{k_0} \text{ P.V.}\int_{-\infty}^\infty \frac{\Pi(\omega')}{\omega - \omega'} d\omega' \nonumber \\
  &= \frac{1}{\pi} \frac{\alpha_0 n_0}{k_0} \ln \left(\frac{\omega_0 - \pi \Gamma -\omega}{\omega_0 + \pi \Gamma - \omega}\right),
\end{align}
where the integration was evaluated using the Hilbert transform for a rectangular function~\cite{king2009}. To determine the group velocity at the center of the spectral transparency window, we observe that
\begin{equation}
  v_g = \frac{c}{n + \omega \frac{dn}{d\omega}},
\end{equation}
where
\begin{equation}
  n = n_0 + \frac{\chi'}{2 n_0}.
\end{equation}
Series expanding $v_g$ around the central frequency $\omega_0$ yields, to first order,
\begin{equation}
  v_g = \frac{\pi^2 \Gamma}{\alpha_0},
  \label{eqn:vg}
\end{equation}
where we have also made use of the assumption that $n \ll \omega \frac{dn}{d\omega}$. Substituting this result for the group velocity into standard Fabry-P\'erot cavity equations yields \eqref{eqn:sl_cav_modes}.


\section{Shot-noise limited power estimate \label{sec:shotnoise_lim}}

The shot-noise limited power for a slow-light cavity is identical to that for a conventional PDH locked cavity. Following Black \cite{black2000}, the spectral density of shot noise has the form
\begin{equation}
  S_e = \sqrt{2 \frac{hc}{\lambda} (2 P_s)},
\end{equation}
and the corresponding frequency noise reads
\begin{equation}
  S_f = \frac{S_e}{\vert D \vert}, 
\end{equation}
where 
\begin{equation}
  D = -\frac{8 \sqrt{P_c P_s}}{\delta \nu_{\text{cav}}}.
\end{equation}
Here $P_c$ and $P_s$ correspond to the carrier and side-band powers, $\nu_{\text{cav}}$ is the cavity linewidth, and the remaining variables have their usual meanings. This yields the following expression for the apparent frequency noise with respect to locking power and the cavity linewidth:
\begin{equation}
  S_f = \frac{1}{4}\sqrt{\frac{h \nu}{P_c}} \delta \nu_{\text{cav}}. 
\end{equation}
For white frequency noise (shot noise) the one-sided noise spectral density $S_y(t)$ has the form
\begin{equation}
  S_y(t) = h_0,
\end{equation}
where $h_0$ is the noise amplitude and can be related to the Allan deviation by \cite{stein1985}
\begin{equation}
  \sigma^2_y(\tau) = \frac{h_0}{2 \tau},
\end{equation}
with $\tau$ the measurement time. Combining the above expressions one obtains the Allan deviation in terms of the locking power
\begin{equation}
  \sigma_y(\tau) = \frac{1}{\sqrt{2\tau}} \frac{1}{4} \sqrt{\frac{h \nu}{P_c}} \delta \nu_{\text{cav}}.
\end{equation}
Considering a locking time of $\tau = 1\;\mathrm{s}$, $\nu=500\;\mathrm{THz}$, $\delta \nu_{\text{cav}} = 1\;\mathrm{kHz}$ and a fractional stability of $\frac{\sigma_y}{\nu} = 1 \times 10^{-17}$, the required locking power corresponds to $P_c = 4.1 \times 10^{-10}\;\mathrm{W}$. As will be demonstrated in appendix~\ref{sec:lockt}, this yields a locking time of more than two hours. This time may be indefinitely extended by utilizing two slow-light cavities and with an interleaved locking and preparation scheme, as has been demonstrated for spectral-hole locking \cite{leibrandt2013}.

\section{Off-resonant excitation width \label{sec:offresex}}

In the section~\ref{sec:discussion} it was noted that when a pulse propagates through a spectral transparency window the surrounding ions off-resonantly absorb photons over a frequency range that scales with the transparency window width $\Gamma$. To arrive at this statement, we begin with a transparency window centered at the origin and, for simplicity, only consider blue detuned frequencies. Further, we note that the off-resonant absorption cross-section fulfills the following proportionality
\begin{equation}
  \sigma(\Delta) \propto \frac{1}{1 + (2 \Delta/\Gamma_h)^2},  
  \label{eqn:abs_crosssec}
\end{equation}
where $\Delta$ is the optical detuning~\cite{allen1987}. We consider the number of photons absorbed by the ions located within an interval of width $\Gamma/2$, half the width of the transparency window, and located immediately adjacent in frequency to the transparency window. Dividing this number by the total number of photons absorbed by all ions positively detuned from the transparency window, one obtains the fraction of absorbed photons
\begin{equation}
  \frac{\int^{\Gamma}_{\Gamma/2} \frac{d\Delta}{1 +(2 \Delta/\Gamma_h)^2}}{\int^{\infty}_{\Gamma/2} \frac{d\Delta}{1 +(2 \Delta/\Gamma_h)^2}} = \frac{\left[\frac{\Gamma_h}{2}\tan^{-1}\left(\frac{2\Delta}{\Gamma_h}\right) \right]^{\Gamma}_{\Gamma/2}}{\left[\frac{\Gamma_h}{2}\tan^{-1}\left(\frac{2\Delta}{\Gamma_h}\right) \right]^{\infty}_{\Gamma/2}} \simeq 0.5,
\end{equation}
where the last equality assumes that $\Gamma/\Gamma_h \gg 1$. An analogous calculation holds for the red detuned ions. Consequently, it follows that 50\% of the off-resonant absorption at the center of the spectral transparency window comes from the atoms in the intervals $[-\Gamma, -\Gamma/2]$ and $[\Gamma/2, \Gamma]$.

\section{Locking time estimate \label{sec:lockt}}

In order to estimate the maximum locking time the spectral transparency window deterioration rate due to off-resonant absorption is required. For an impedance matched cavity, all of the photons in the carrier can be assumed to be absorbed. For a locking power of $P_c = 4.1 \times 10^{-10}\;\mathrm{W}$ the photon flux is 
\begin{equation}
  N = \frac{P_c}{h \nu} \approx 1.3 \times 10^{9}/\mathrm{s}.  
\end{equation}
Considering only the ions that are located within a frequency width $\Gamma$ on either side of the transparency window, this corresponds to half of the total number of photons absorbed (shown in appendix \ref{sec:offresex}). Defining this as $N_{\mathrm{abs},\Gamma} = N/2 = 6.5 \times 10^{8}/\mathrm{s}$, the locking time can then be estimated as
\begin{equation}
  T = \frac{N_{\mathrm{ions}}}{N_{\mathrm{abs}, \Gamma}},
\end{equation}
where $N_{\mathrm{ions}}$ is the total number of ions in the interaction volume that can be excited before the spectral transparency window shape deteriorates to a level that makes locking unfeasible. For an inhomogeneous linewidth of $\Gamma_{\mathrm{inhom}} = 22\;\mathrm{GHz}$, the dopant ion density has the form
\begin{equation}
  \rho_{\mathrm{ions}} = \frac{\rho_{\mathrm{host}} \delta}{\Gamma_{\mathrm{inhom}}} \approx 4 \times 10^{15}/(\mathrm{m}^3\mathrm{Hz}).
\end{equation}
Here, $\rho_{\mathrm{host}} = 9.15 \times 10^{27}/\mathrm{m}^3$ is the density of yttrium atoms located at site 1 for which the dopant can substitute in Y$_2$SiO$_5$ (assuming equal site 1 and site 2 occupation) \cite{konz2003}, and $\delta = 1\%$ corresponds to the dopant concentration. As photons scatter off ions that are adjacent in frequency to the transparency window, there is a probability that the ions will be pumped to a different state which would broaden the cavity linewidth and therefore require an increase in the locking power in order to maintain the same Allan deviation. Somewhat arbitrarily setting the threshold where this starts to affect the locking at 10 \%, one obtains 
\begin{equation}
  N_{\mathrm{ions}} = 0.1 \rho_{\mathrm{ions}} \Gamma V = 5 \times 10^{12},
\end{equation}
where the interaction volume $V$ was taken as $0.25 \times 10^{-6}\;\mathrm{m}^3$. This leads to a total locking time of $128\;\mathrm{min}$. Note that we have neglected the probability that ions that are excited decay back to the same state from which they were excited. For the branching ratios of the relevant transition in Eu$^{3}$:Y$_2$SiO$_5$ there is a significant probability this can occur; however, including this effect would only improve the locking time. Consequently, it is perfectly reasonable to ignore this effect in order to obtain a lower bound on the attainable locking time.

\section{Data analysis details \label{sec:data_analysis}}

\begin{figure}[ht!]
\centering
\includegraphics[width=0.6\columnwidth]{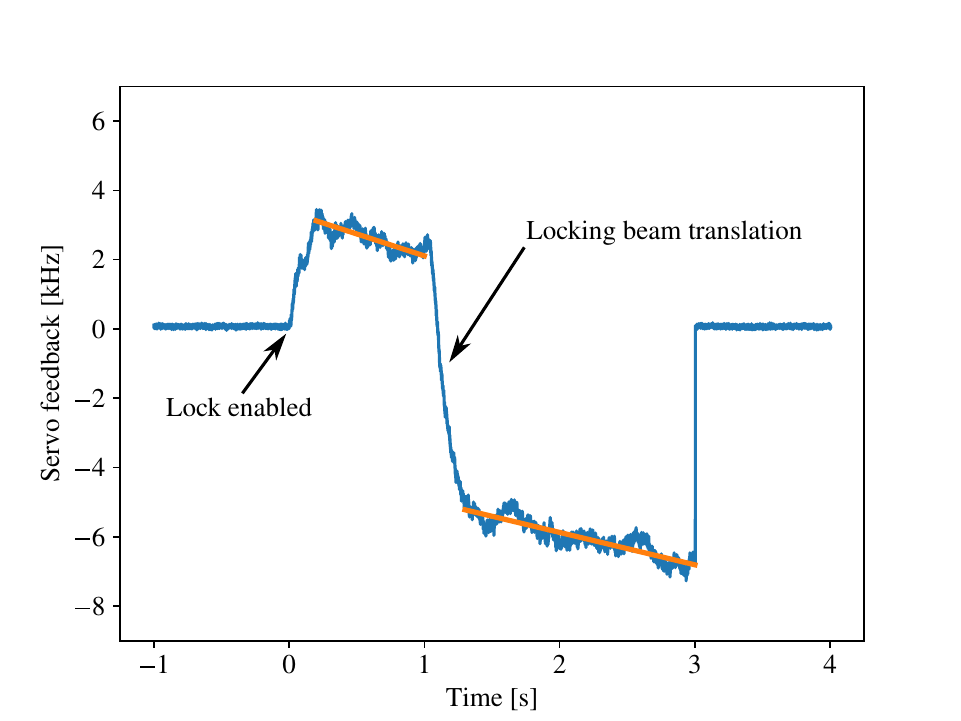}
\caption{The servo feedback recorded while the locking beam was translated horizontally by $2\;\mathrm{\mu m}$. The lock was enabled at $t=0\;\mathrm{s}$; the large initial feedback signal is attributed to an offset of the initial locking point, as well as a small amount of unburned ions near the mirror coatings leading to a local perturbation of the dispersion. This latter effect was largely mitigated by diffusing and dithering the perpendicular burning beam, as described in the methods section. The locking beam translation occurs $2\;\mathrm{\mu m}$ at $t=1\;\mathrm{s}$.}
\label{fig:twoumstep}
\end{figure}

In this section we discuss details of the data analysis performed to produce key figures in the main text. In particular, figure \ref{fig:twoumstep} shows the frequency feedback signal recorded as the length of the locking cavity was altered by translating the beam along the wedged surface of the locking crystal. The trace corresponds to the PI servo output monitored via a high-impedance buffer circuit. 

The plot shows a steep initial ramp, which is attributed to residual ions near the mirror coatings due to a standing wave pattern set up by the perpendicular burning beam. This effect was minimized by dithering the perpendicular burning beam, but some residual ions remained. For a practical implementation of this locking scheme, a material with a slower spin-lattice relaxation rate would resolve this issue. A linear function was fitted to the feedback both before and after the step; this allowed for any residual drift to be subtracted. The time required to complete a given step size was determined by inspection of representative servo-feedback data. The frequency change error bars in figure~\ref{fig:cav_modes} were determined by evaluating the standard error of 10 independent measurements. However, there could be a systematic error in the step timing interval selection, which would be missed by this error analysis. The error bars for the cavity length change also assumed that errors were randomly distributed; if there was a systematic error in the stepper motor, this would not be accounted for in our uncertainty analysis. We note that effects of backlash for the motor driven mirror were mitigated using a servo readout. 

\begin{figure}[tb!]
\centering
\includegraphics[width=0.6\columnwidth]{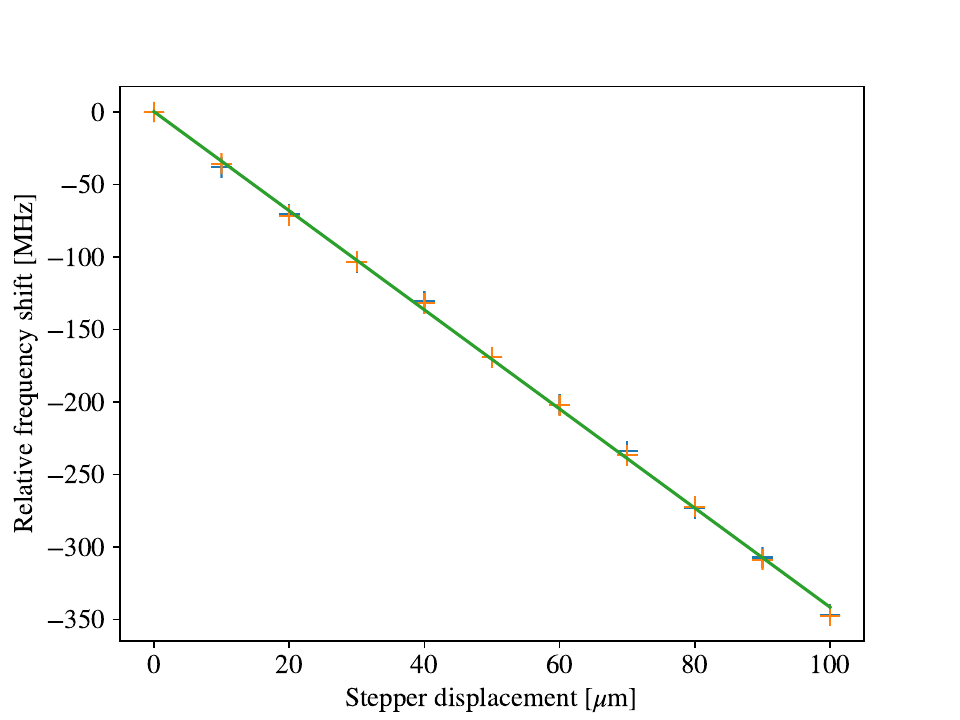}
\caption{The frequency shift with respect to the locking beam displacement when detuned by $\simeq 5.3\;\mathrm{THz}$ from the Pr$^{3+}$ ensemble. The frequency shifts were obtained from the resonance frequencies of the cavity observed in transmission for each stepper motor displacement.}
\label{fig:offres_shift}
\end{figure}

The above outlined procedure was employed for analyzing the beam translation experiments for the group velocities of $v_g = c/7500$ and $v_g = c/2790$. In the scenario where there was no slow-light effect, the frequency shift data was obtained by translating the beam over a range of steps and performing a linear fit to the cavity mode peaks (determined from a Lorentzian fit). This data is summarized in figure~\ref{fig:offres_shift}. The uncertainties in the frequency shift were obtained from the covariance matrix of the first order polynomial fit shown in figure~\ref{fig:offres_shift}. 

\begin{figure}[tbh!]
\centering
\includegraphics[width=0.6\columnwidth]{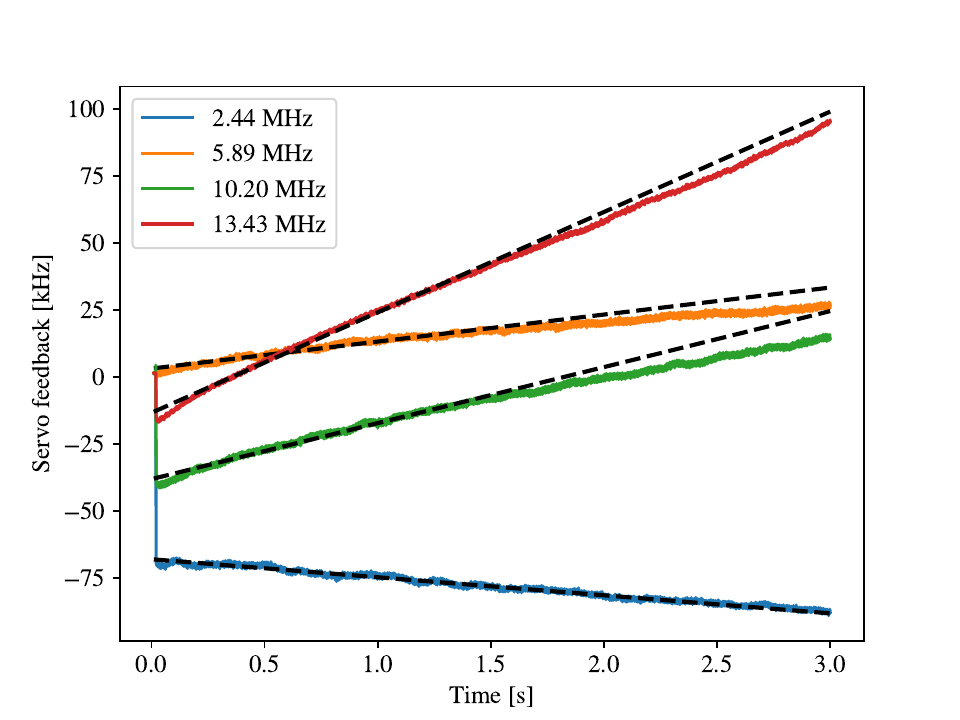}
\caption{Representative locking feedback data when the laser was locked to different cavity modes across the spectral transparency window. The different slopes originate from the varying rates of off-resonant excitation combined with the fast hyperfine cross-relaxation between the $\ket{\pm1/2g}$ and $\ket{\pm 3/2g}$ states. Comparable data was obtained for a range of optical locking power intensities.}
\label{fig:drift_traces}
\end{figure}

Figure~\ref{fig:drift_traces} shows the drift data from which the $6\;\mathrm{\mu W}$ data points of figure~\ref{fig:drift_wrt_power} were obtained. These are representative of the drift data obtained for other locking beam powers. The indicated frequencies correspond to the cavity mode center frequencies to which the laser was locked. A linear fit was performed using the frequency interval between $0.1-1.5\;\mathrm{s}$; this restricted fitting interval was selected since, as discussed above, for larger locking beam powers off-resonant excitation lead to a non-linear drift at longer locking times. The error bars shown in figure~\ref{fig:drift_wrt_power} were determined from the square roots of the corresponding covariance matrices.

%

\end{document}